\documentclass[a4paper,11pt]{article}
\usepackage{amsmath, amsthm, amssymb,pdflscape,bm,rotating,times,color,longtable,url}
\usepackage{graphicx}
\usepackage{multirow}

\usepackage{psfrag}
\usepackage[round]{natbib}
\usepackage[T1]{fontenc}
\usepackage[sc]{mathpazo}
\begin{document}

\title{An integer-valued time series model for multivariate surveillance}
\author{Xanthi Pedeli \\
Department of Environmental Sciences, Informatics \& Statistics
\\Ca' Foscari University of Venice, \\Italy
 \and Dimitris Karlis  \\ Department of Statistics, \\
Athens University of Economics \\ Greece}

\date{}
\maketitle

\begin{abstract}
In recent days different types of surveillance data are becoming available for public health reasons.  In most cases several variables are monitored and events of different types are
reported. As the amount of surveillance data increases, statistical methods that can effectively address multivariate surveillance scenarios are demanded. Even though research activity
in this field is increasing rapidly in recent years, only a few approaches have simultaneously addressed the integer-valued property of the data and its correlation (both time correlation
and cross correlation) structure. In this paper, we suggest a multivariate integer-valued autoregressive model that allows for both serial and cross correlation between the series and can
easily accommodate overdispersion and covariate information. Moreover, its structure implies a natural decomposition into an endemic and an epidemic component, a common distinction in dynamic models for infectious disease counts. Detection of disease outbreaks is achieved through the comparison of surveillance data with one-step-ahead predictions obtained after fitting the suggested model to a set of clean historical data. The performance of the suggested model is illustrated on a trivariate series of syndromic surveillance data collected during Athens 2004 Olympic Games.

\end{abstract}

\section{Introduction}
\label{intro}

The main purpose of public health surveillance systems is the
effective and timely detection of a disease outbreak so that
control measures for the elimination of disease transmission can
be rapidly taken.  Over the last two decades, several statistical
techniques such as statistical process control methods and
statistical modelling techniques have been developed to this
direction \cite{sonesson2003,
shmueli2010, unkel2012}. These methods aim to identify unusual
patterns in data series which may result from infectious disease
outbreaks. Particular features that characterize these data series
actually pose the statistical challenges in this research area.

More specifically, health data typically consist of non-negative counts representing numbers of events
at specific time points. Consider for example the daily number of diagnosed cases or emergency department
visits. In such cases, the normality assumption, that is fundamental for the construction of traditional control
charts, is rather inappropriate and should be replaced by other, more suitable assumptions, like those of Poisson
or negative binomial distributed data.
The Poisson CUSUM \cite{lucas1985} and Poisson EWMA \cite{borror1998} charts consist early adaptations of control charts for count data.
Regression models
with non-normal errors have also been suggested in the syndromic surveillance literature. Examples include
Poisson log-linear models \cite{parker1989, farrington1996, jackson2007} and the count data regression charts constructed under the assumption of a generalized linear model for negative binomial counts \cite{hohle2008}.

Another issue concerning data used in public health surveillance is that health data are usually
correlated over time. Failure to account properly for autocorrelation can result in a misspecified model,
introducing bias in the estimated effects and prediction intervals. In order to capture the correlation structure
of health data, time series techniques have been used in the field of public health surveillance and outbreak detection.
More often, autoregressive integrated moving average (ARIMA) models have been employed to model the variation of
health data and their dependence on past observations \cite{choi1981, helfenstein1986, watier1991, reis2003}. More sophisticated approaches as for example Bayesian and Hidden Markov models have also been
suggested \cite{lestrat1999, held2006, hohle2007, hohle2010}. Extensive accounts of
time series models and their limitations  in public health surveillance (e.g. non-stationarity of surveillance data)
can be found in \cite{unkel2012} and \cite{siripanthana2013}.

Generally, monitoring of a single series is of limited interest for outbreak detection purposes. For the early detection
of a large-scale bioterrorism or epidemic outbreak, multivariate data should be considered. Such data can be a specific
variable measured in several regions, or different variables measured in a particular region, or multiple variables measured
in multiple regions \citep{siripanthana2013}. In all these situations, the same underlying process may have generated several
data series introducing correlations between them. To account for such correlations, and hence improve the timeliness in outbreak
detection, multivariate surveillance has been proposed. Statistical methods used in multivariate surveillance are categorized into the following groups \cite{sonesson2005, frisen2010}: dimensionality reduction (including principal components
and sufficient reduction techniques), parallel surveillance (where each series is monitored separately by means of a univariate
surveillance method), joint modeling (with alarm functions based on the likelihood ratio statistic), scalar accumulation
(Hotelling's $T^2$ charts) and vector accumulation methods (MCUSUM and MEWMA charts).
For multivariate counts \cite{held2017probabilistic} use an observation driven model for surveillance,
see also the work of \cite{maelle2014monitoring} for related material. The approach of
\cite{schioler2012multivariate} for continuous outcomes is also relevant.

Even though research activity in the field of multivariate surveillance increases rapidly in recent years, only a few
approaches have simultaneously addressed the integer-valued property of the data and its correlation structure.
Closest to such a comprehensive approach are the works of  \cite{held2005, paul2008} and \cite{paul2011} who use multivariate
branching processes for modeling multivariate infectious disease surveillance data. In this paper, we suggest a different
modeling approach based on the class of integer-valued autoregressive (INAR) processes.
INAR models have been used in the context of health surveillance and outbreak detection by a few authors in the past.
In a paper by \cite{cardinal1999} INAR and ARIMA models are both applied for the analysis of infectious disease incidence and it is shown that
the relative forecast error is smaller in the first instance. In a statistical process control framework, \cite{weiss2009, weiss2011}
and \cite{weisstestik2009} proposed EWMA and CUSUM charts respectively for Poisson INAR(1) processes.
For monitoring multivariate Poisson counts  one can see the works of \cite{CozzucoliQTQM2018} and  \cite{wang2017statistical}.

In the context of multivariate health surveillance, a sufficient
reduction method has been suggested  \cite{siripanthana2013} for the detection of a shift in a bivariate Poisson INAR(1) process \citep{pedeli2011}.
We take another perspective where a multivariate INAR(1) model is fitted to health data in order to make a
prediction as a threshold for the next count. We consider a simplified version of the multivariate INAR(1) process
proposed by \cite{pedeli2013} where the innovations $\epsilon_t$ are assumed to be independent random variables.
The introduced process has several merits. Firstly, it admits a realistic epidemiological interpretation especially
for spatial and syndromic surveillance. Secondly, it accounts for the relationship with time lag between series,
which is typical in disease transmission. Finally, it can easily accommodate overdispersion and covariate information
and is extremely advantageous in terms of practical implementation.

The rest of the paper is organized as follows. Section 2  summarizes main concepts on multivariate integer-valued autoregressive processes and outlines the model and its properties.
The suggested methodology is described in Section 3. Section 4 includes a simulation study for the evaluation of the suggested outbreak detection statistical process. A multivariate series of syndromic surveillance data
illustrates our approach in Section 5. Section 6 concludes the paper.

\section{Multivariate integer--valued autoregressive models}

Integer--valued autoregressive (INAR) models have been introduced in the statistical literature by \cite{mckenzie1985} and \cite{alosh1987}
as discrete counterparts of the standard Gaussian autoregressive process. The simplest integer-valued autoregressive model is that of order one, denoted briefly as INAR(1) and defined as
$X_{t}=\alpha\circ X_{t-1}+\epsilon_{t}$, $t\in\mathbb{N},$
where $\{\epsilon_{t}\}$ is an innovation process consisting of uncorrelated non--negative integer--valued random variables with finite mean and variance. The $\circ$ symbol denotes the binomial thinning operator defined by $\alpha\circ X=\sum_{j=1}^X Y_j$ where $\{Y_j\}$ are independent identically distributed Bernoulli random variables with $P(Y_j=1)=1-P(Y_j=0)=\alpha$. The binomial thinning operator preserves the integer nature of the INAR process and introduce serial dependence though conditioning on $X_{t-1}$.
 In recent years, the wide interest in the class of INAR models has resulted in its enrichment with several variants and extensions of the benchark  integer-valued autoregressive model of order one. We concentrate on multivariate extensions of the INAR(1) process as considered by \cite{pedeli2013} and \cite{pedeli2013b} and briefly summarized in the following.

\paragraph{Full multivariate INAR(1) model}
Assume that $\bm{A}$ is a $n \times n$ matrix with entries $\alpha_{ij}$ satisfying
$0\leq\alpha_{ij}\leq 1$ for $i,j=1,\ldots,n$ and let $\bm{X}$ be a random vector with values in $\mathbb{N}^{n}$.
Then, $\bm{A}\circ\bm{X}$ is a $n$-dimensional random vector with $i$-th component
\begin{equation}
\label{multithin}
[\bm{A}\circ\bm{X}]_{i} = \sum_{j=1}^{n}\alpha_{ij}\circ X_{j},\quad i=1,\ldots,n
\end{equation}
where the counting series in all $\alpha_{ij}\circ X_j$, $i,j=1,\ldots,n$ are assumed to be independent.
Based on (\ref{multithin}) we can define a full multivariate integer-valued autoregressive process of order 1 (MINAR(1)) \citep{pedeli2013}
\begin{equation}
\label{minar1}
\bm{X}_t=\bm{A}\circ\bm{X}_{t-1}+\bm{\epsilon}_t, \quad t\in\mathbb{Z},
\end{equation}
where $\{\bm{\epsilon}_t\}_{t\in\mathbb{Z}}$ is a sequence of non-negative integer-valued
random vectors, with mean $\bm{\mu}_{\bm{\epsilon}}$ and variance matrix $\bm{\Sigma}_{\bm{\epsilon}}$,
independent of $\bm{A}\circ\bm{X}_{t-1}$.
Therefore, the $i$th element of the full MINAR(1) process is given by
$X_{it} = \sum_{j=1}^{n}\alpha_{ij}\circ X_{j,t-1} + \epsilon_{it}$, $i=1,\ldots,n$,
where $\alpha_{ij}\circ X$ are assumed to be mutually independent binomial thinning operations. 
The non-negative integer-valued random process $\{\bm{X}_t\}_{t\in\mathbb{Z}}$ is the unique
strictly stationary solution of (\ref{minar1}) if the largest eigenvalue of $\bm{A}$ is less than 1 and
$E||\bm{\epsilon}_t||<\infty$.

Using basic thinning operation properties \citep{franke1993, latour1997}, it can be shown that the mean vector
and variance covariance matrix of the process $\bm{X}_t$ are given by
\begin{equation}
\bm{\mu} = E(\bm{X}_t) = (\bm{I}-\bm{A})^{-1}\bm{\mu}_{\bm{\epsilon}},
\label{minar-mu}
\end{equation}
and
\begin{eqnarray}
\label{minar-var}
\bm{\gamma}(h) &=& E[(\bm{X}_{t+h} - \bm{\mu})(\bm{X}_t - \bm{\mu})^T] \nonumber\\
&=&\left\{\begin{array}{ll}
\bm{A}\bm{\gamma}(0)\bm{A}^T+\mbox{diag}(\bm{B}\bm{\mu}) + \bm{\Sigma}_{\bm{\epsilon}}, & h= 0\\
\bm{A}^{h}\bm{\gamma}(0), & h\geq 1
\end{array}\right.
\end{eqnarray}
respectively, where $[\bm{B}]_{ij} = \alpha_{ij}(1-\alpha_{ij})$ for $i,j = 1,\ldots,n$ \citep{pedeli2013}.

The unknown parameter $\bm{\theta}=(\mbox{vec}(\bm{A})^{'},\bm{\mu_{\epsilon}}^{'},\mbox{vec}(\bm{\Sigma_{\epsilon}})^{'})^{'}$ of the full MINAR(1) process can be estimated using the method of conditional maximum likelihood \citep{pedeli2013}.
The maximum likelihood estimator (MLE) of $\bm{\theta}$ is defined as $\hat{\bm{\theta}}=\mbox{argmax}_{\bm{\theta}}\ell(\bm{\theta})$
where
\begin{equation}
\label{minarloglik}
\ell(\bm{\theta})=\sum_{t=2}^{T}\log f(\mathbf{x}_{t}|\mathbf{x}_{t-1},\boldsymbol{\theta}),
\end{equation}
is the conditional log likelihood function and $T$ is the length of the time series. The conditional densities involved in the right-hand side of (\ref{minarloglik})
are convolutions of $n$ sums of binomials
$f_{i}(x_{i}|\mathbf{x}_{t-1})=P(X_{it}=x_{i}|\mathbf{X}_{t-1}=\mathbf{x}_{t-1}), \quad i=1,\ldots,n,$
and a distribution of the form
$g(k_{1},\ldots,k_{n})=P(\epsilon_{1t}=k_{1},\ldots,\epsilon_{nt}=k_{n}),$
corresponding to the joint distribution of the innovations $\{\bm{\epsilon}_t\}$. Hence, $f(\mathbf{x}_{t}|\mathbf{x}_{t-1},\boldsymbol{\theta})$
can be expressed as the multiple sum
\begin{equation*}
f(\mathbf{x}_{t}|\mathbf{x}_{t-1},\boldsymbol{\theta})=\sum_{k_1=0}^{m_1}\cdots\sum_{k_n=0}^{m_n}f_{1}(x_{1t}-k_{1}|\mathbf{x}_{t-1})\cdots f_{n}(x_{nt}-k_{n}|\mathbf{x}_{t-1})g(k_{1},\ldots,k_{n}),
\end{equation*}
where $m_{i}=\mbox{min}(x_{it},x_{i;t-1})$, $i=1,\ldots,n$.

\paragraph{Constrained multivariate INAR(1) model}
For the maximization of (\ref{minarloglik}) one can resort to numerical techniques. However, under the assumption of a cross-correlated innovation process,
the complexity of (\ref{minar1}) and hence the numerical difficulty of the maximum likelihood approach, increase sharply with dimensional increase
\citep{pedeli2013}. To avoid such complications, \cite{pedeli2013b} consider a constrained MINAR(1) model by assuming a single source of dependence between
the univariate series that comprise the MINAR(1) process. In particular, $\bm{A}$ is assumed to be a $n\times n$ diagonal matrix with independent
elements $\alpha_{i}=[\bm{A}]_{ii}$, $i=1,\ldots,n$, while $\{\bm{\epsilon}_t\}$ follow jointly a discrete
multivariate distribution.
One can easily see that the assumption of a diagonal $\bm{A}$ matrix significantly reduces the correlation structure since
each univariate series $\{X_{it}\}$ at time $t$ is a function of each own predecessors at time $t-1$ but not of the predecessors of the rest of the series
comprising the MINAR(1) process, i.e.
$X_{it}=\alpha_{i}\circ X_{i,t-1} + \epsilon_{it}, \quad i=2,\ldots,n.$
For the estimation of this simplified model, \cite{pedeli2013b} suggest a pairwise likelihood approach that reduces the multivariate estimation problem to a
set of bivariate problems. 

\section{Surveillance using a new MINAR(1) specification}
\label{sec-survmodel}
Traditionally, statistical models for health surveillance data aim to effectively capture the endemic and epidemic dynamics of disease risk. In principle, the endemic component explains a baseline rate of cases with stable temporal pattern. More specifically, it describes the risk of new events as a function of external factors independent of the history of the epidemic process. Seasonality, socio-demographic characteristics, population density and vaccination coverage are a few examples of such factors. The epidemic component on the other hand aims to introduce infectiousness, that is explicit dependence between events. Therefore the epidemic component is driven by the observed past and is identified with the autoregressive part of the model \citep{meyer2016}.

This additive decomposition of disease risk is well embodied in model (\ref{minar1}). However, the simplification of (\ref{minar1}) based on the assumption of a diagonal matrix $\mathbf{A}$ weakens the ability of the model to capture the epidemic dynamics of disease risk since it ignores the relationship with time lag between series that is typical in disease transmission. Moreover, inference for the constrained multivariate INAR(1) model is based on a pairwise likelihood approach \cite{pedeli2013b} which is not appropriate for prediction purposes since 
the accuracy of the predictive conclusions is affected by the fact that they are based on a misspecified
model (misspecification error).

To balance between effectiveness and attractiveness of the model, we consider here another simplification of (\ref{minar1}). In particular, we assume that the correlation matrix $\mathbf{A}$ is non-diagonal and we relax the degree of complexity of the model by assuming that the innovation series $\boldsymbol{\epsilon}_t$, i.e. the endemic components, are uncorrelated. The resulting model admits a realistic epidemiological interpretation and is extremely advantageous in terms of practical implementation since the distribution of the innovations becomes a product of univariate mass functions, i.e.
\begin{equation*}
g(k_{1},\ldots,k_{n})=\prod_{i=1}^{n}P(\epsilon_{it}=k_{i}).
\end{equation*}
The mean vector and variance covariance matrix of the new MINAR(1) process $\bm{X}_t$ are still given by (\ref{minar-mu}) and (\ref{minar-var}) respectively where $\bm{\Sigma}_{\bm{\epsilon}}$ is now a diagonal matrix.

Moreover, overdispersion that is a typical characteristic of health surveillance data, can be easily accommodated even under the simplest parametric assumption of Poisson innovations. More specifically, in line with \cite{pedeli2013b} it can be proved that if $\boldsymbol{\epsilon_{t}} = (\epsilon_{1t},\ldots,\epsilon_{nt})^{'}$ are independent Poisson random variables with parameters $\boldsymbol{\lambda} = (\lambda_1,\ldots,\lambda_n)^{'}$, then the joint distribution of $\mathbf{X}_{t}$ is given by the product of $n$ generalized Poisson distributions \cite{gurland1957, kemp1965} with parameters that are nonlinear combinations of
$\boldsymbol{\lambda}$ and powers of $\mathbf{A}$.
In the bivariate case ($n=2$), it is straightforward to show that the vector of expectations $\boldsymbol{\mu} = (\mu_1, \mu_2)^{'}$ and the variance covariance matrix $\boldsymbol{\gamma}(0)=\left[\begin{array}{cc}\gamma_{11}(0) & \gamma_{12}(0)\\
\gamma_{12}(0) & \gamma_{22}(0)
\end{array}\right]$ have elements
\begin{equation*}
\mu_1 = \frac{(1-\alpha_{22})\lambda_1 + \alpha_{12}\lambda_2}{(1-\alpha_{11})(1-\alpha_{22})-\alpha_{12}\alpha_{21}}, \quad
\mu_2 = \frac{(1-\alpha_{11})\lambda_2 + \alpha_{21}\lambda_1}{(1-\alpha_{11})(1-\alpha_{22})-\alpha_{12}\alpha_{21}},
\end{equation*}
and
\begin{eqnarray*}
\gamma_{11}(0) & = & \frac{(1-\alpha_{11}^2)\mu_{1}+\alpha_{12}^{2}(\gamma_{22}(0)-\mu_{2})+2\alpha_{11}\alpha_{12}\gamma_{12}(0)}{1-\alpha_{11}^2},\\
\gamma_{22}(0) & = & \frac{(1-\alpha_{22}^2)\mu_{2}+\alpha_{21}^{2}(\gamma_{11}(0)-\mu_{1})+2\alpha_{22}\alpha_{21}\gamma_{12}(0)}{1-\alpha_{22}^2},\\
\gamma_{12}(0)&=&\frac{\alpha_{11}\alpha_{21}\gamma_{11}(0)+\alpha_{22}\alpha_{12}\gamma_{22}(0)}{1-\alpha_{11}\alpha_{22}-\alpha_{12}\alpha_{21}},
\end{eqnarray*}
respectively.
For higher dimensions, we confine to the gereral formulae (\ref{minar-mu}) and (\ref{minar-var}) since closed form expressions for the elements of $\boldsymbol{\mu}$ and $\boldsymbol{\gamma}(0)$ cannot be easily derived. Other parametric families, as e.g. the negative binomial distribution, can also be easily considered for the distribution of the innovations depending on the degree of overdispersion present in the data.

Conditional maximum likelihood estimates for the new MINAR(1) process can be obtained through maximization of the likelihood function
\begin{equation*}
L(\boldsymbol{\theta}|\mathbf{x})=\prod_{t=2}^{T}f(\mathbf{x}_{t}|\mathbf{x}_{t-1},\boldsymbol{\theta}),
\end{equation*}
where $f(\mathbf{x}_{t}|\mathbf{x}_{t-1},\boldsymbol{\theta})$ is the conditional density of $\boldsymbol{X}_t$ given $\boldsymbol{X}_{t-1}$ and $\boldsymbol{\theta}$ is the vector of unknown parameters.
For instance, in the bivariate case and under the assumption of Poisson innovations,
\begin{eqnarray*}
\label{modifcdfPois}
&&f(\mathbf{x}_{t}|\mathbf{x}_{t-1},\boldsymbol{\theta})=
\sum_{k_1=0}^{\min(x_{1t},x_{1,t-1}, x_{2,t-1})}\sum_{k_2=0}^{\min(x_{2t},x_{1,t-1}, x_{2,t-1})}\left\{ e^{-(\lambda_{1}+\lambda_{2})}\sum_{m=0}^{\min(k_1,k_2)}\frac{\lambda_{1}^{x_{1t}-k_1-m}\lambda_{2}^{x_{2t}-k_2-m}}{(x_{1t}-k_1-m)!(x_{2t}-k_2-m)!}\right. \nonumber\\
&&\times
\sum_{j_{1}=0}^{k_1}\left(
\begin{array}{c}
x_{1,t-1}\\
j_{1}\\
\end{array}\right)\left(
\begin{array}{c}
x_{2,t-1}\\
k_1-j_{1}\\
\end{array}\right)
\alpha_{11}^{j_{1}}(1-\alpha_{11})^{x_{1,t-1}-j_{1}}
\alpha_{12}^{k_1-j_{1}}(1-\alpha_{12})^{x_{2,t-1}-k_1+j_{1}}
\nonumber\\
&& \left.\times\sum_{j_{2}=0}^{k_2}\left(
\begin{array}{c}
x_{2,t-1}\\
j_{2}\\
\end{array}\right)\left(
\begin{array}{c}
x_{1,t-1}\\
k_2-j_{2}\\
\end{array}\right)
\alpha_{22}^{j_{2}}(1-\alpha_{22})^{x_{2,t-1}-j_{2}}
\alpha_{21}^{k_2-j_{2}}(1-\alpha_{21})^{x_{1,t-1}-k_2+j_{2}}\right\} \nonumber\\
\end{eqnarray*}
where $\boldsymbol{\theta}=\{\alpha_{11},\alpha_{12},\alpha_{21},\alpha_{22},\lambda_{1},\lambda_{2}\}$.
If other parametric assumptions are made as e.g. negative binomial innovations, the above formula can be modified accordingly.

The newly defined multivariate INAR(1) process can be used for modeling clean historical data and make one-step-ahead forecasts that can be used for prediction-based monitoring. We should emphasize here that the set-up phase is assumed to be free or cleaned of outbreaks.
The suggested outbreak detection statistical process comprises of two steps: In the first step, the available series of data in the set-up phase (historical data) is modeled
through a multivariate INAR(1) process and a parameter vector of maximum likelihood estimates $\hat{\boldsymbol{\theta}}$ is obtained. The second step is dedicated to the successive monitoring of incoming observations in the operational phase (surveillance data) using the model obtained from the set-up phase. In particular, the actually observed realization $\mathbf{x}_{t+1}$ is assessed against a multivariate prediction threshold derived from the model fitted in the first step in order to define whether an alarm should be triggered.
More specifically, for each multivariate observation $\mathbf{x}_{t+1}$ in the operational phase, we estimate  the one-step-ahead predictive distribution
$\hat{P}(\mathbf{X}_{t+1}=\mathbf{x}_{t+1}|\mathbf{x}_{t},\hat{\boldsymbol{\theta}})$,
$\mathbf{x}\in\mathbb{N}_{0}^{n}$ and obtain the marginal predictive probabilities $\hat{P}(X_{i,t+1}=x_{i,t+1}|\mathbf{x}_{t},\hat{\boldsymbol{\theta}})$, $i=1,\ldots, n$.
For each observation $X_{i,t+1}$, we construct an $(1-\alpha)\%$ prediction interval with upper bound $x^{UB}_{i,t+1}$ equal to the $(1-\alpha)$-quantile of the corresponding marginal predictive distribution, where $\alpha$ is a prespecified significance level. The lower bound of the prediction interval is set equal to 0 since we are only interested in detecting positive deviations from the in-control model. Each series flags an alarm at time $t+1$ if the corresponding observation lies outside the prediction interval, i.e. if
$$x_{i,t+1}>x_{i,t+1}^{UB}.$$
Finally, for the overall alarm, a majority rule can be defined, i.e. flagging an alarm if a certain percentage of the series signals an alarm at the same point in time \citep{vial2016}.

\section{Simulation study}
\label{sec:simulations}

We conducted a small simulation study aiming to evaluate the
performance of the suggested outbreak detection statistical
process. Time series data of length $n=200$ were simulated from a
trivariate INAR(1) model with independent Poisson innovations. We
assumed that the first $150$ observations consist the set-up phase
(that is a clean process without outbreaks) and the last $50$
observations consist the monitoring phase. Subsequently, for each
series $i$, $i=1,2,3$, we simulated an outbreak  of expected size
$\kappa_i$ at time $t=170$ from a Poisson distribution with mean
equal to $\kappa_i$. Therefore, our model has the form
$$\left(\begin{array}{c}
X_{1t}\\X_{2t}\\X_{3t}\end{array}\right) =
\left[\begin{array}{ccc}
\alpha_{11} & \alpha_{12} & \alpha_{13}\\
\alpha_{21} & \alpha_{22} & \alpha_{23}\\
\alpha_{31} & \alpha_{32} & \alpha_{33}\\
\end{array}\right]\circ
\left(\begin{array}{c}
X_{1,t-1}\\X_{2,t-1}\\X_{3,t-1}\end{array}\right)+
\left(\begin{array}{c}
\epsilon_{1t}\\\epsilon_{2t}\\\epsilon_{3t}\end{array}\right),
$$
where $\epsilon_{it}$ are independent Poisson random variables
with mean $E(\epsilon_{it})=\lambda_i + \kappa_i I(t=170)$ and
$I(A)$ is an indicator function.

\begin{figure}
\centering
\includegraphics[scale=0.55]{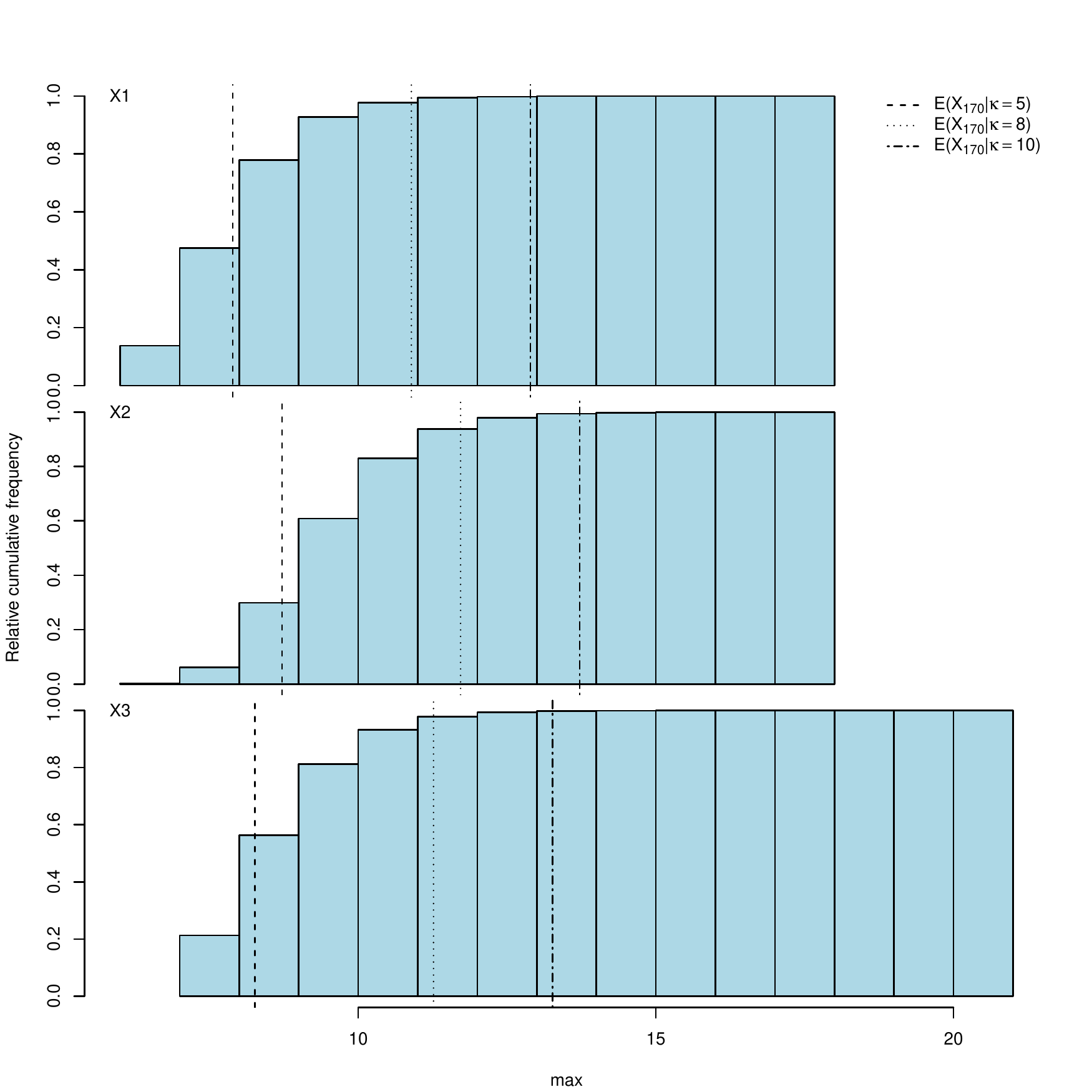}
\caption{Cumulative distribution of the maximum values of $10000$ triavariate INAR(1) series with independent
Poisson innovations. Data have been simulated with $n=200$ and $\mathbf{\theta}=(\alpha_{11},\alpha_{12},\alpha_{13},\alpha_{21},\alpha_{22},\alpha_{23},\alpha_{31},\alpha_{32},\alpha_{33},\lambda_1,\lambda_2,$ $\lambda_3)= (0.3, 0.1,0.2,0.2,0.4,0.2,0.3,0.2,0.2,1,1,1)$.}
\label{kappa-fig}
\end{figure}

The true parameter values were assumed to be
$$\boldsymbol{A}=\left[\begin{array}{ccc}
\alpha_{11} & \alpha_{12} & \alpha_{13}\\
\alpha_{21} & \alpha_{22} & \alpha_{23}\\
\alpha_{31} & \alpha_{32} & \alpha_{33}\\
\end{array}\right] =
\left[\begin{array}{ccc} 0.3 & 0.1 & 0.2\\
0.2 & 0.4 & 0.2\\
 0.3 & 0.2 & 0.2 \\
\end{array}\right],$$
and $\lambda_1=\lambda_2=\lambda_3=1$. We also assumed that
$\kappa_1=\kappa_2=\kappa_3=\kappa$ and we took the values of
$\kappa$ to be 5, 8 or 10. 
By choosing these specific values for $\kappa$ we aim to study both cases where the outbreak is 
manifest, as well as cases where the outbreak cannot be easily distinguished from the typical
range of values in the in-control state.
Figure~\ref{kappa-fig} 
shows the cumulative distribution of the maximum values of $10000$ trivariate INAR(1) series with independent Poisson innovations and $n=200$ observations.
All trivariate series are free of outbreaks ($\kappa=0$) and have been simulated with parameter values as described above.

The maximum values range between 6 and 18 for $\{X_1\}$ and $\{X_2\}$ and between 7 and 21 for $\{X_3\}$.
Since the process is in contol until $t=169$, the expected value of the trivariate series at the time of the outbreak ($t=170$) can be easily computed as $E(\boldsymbol{X}_{170})=\boldsymbol{A}E(\boldsymbol{X}_{169})+\boldsymbol{\lambda}+\kappa=\boldsymbol{A}(\boldsymbol{I}-\boldsymbol{A})^{-1}\boldsymbol{\lambda}+\boldsymbol{\lambda}+\kappa$, where $\boldsymbol{\lambda}=(\lambda_1,\lambda_2,\lambda_3)^{'}$. The computed expectations for $\kappa=5, 8$ and $10$ are summarized in Table~\ref{tab:kappa-x} and illustrated in Figure~\ref{kappa-fig} with vertical lines. Table~\ref{tab:kappa-xmax} summarizes the empirical probabilities of the maximum value of each univariate series being greater than the corresponding expectation of the series at the time of an outbreak ($t=170$), $P(\underset{t\neq 170}{\max}(x_{it})>E(X_{i,170}))$, $i=1,2,3$. From Figure~\ref{kappa-fig} and Table~\ref{tab:kappa-xmax} we can conclude that $\kappa=5$ corresponds to outbreaks that cannot be easily distinguished from the typical range of values in the in-control state, since the empirical probabilities $P(\underset{t\neq 170}{\max}(x_{it})>E(X_{i,170}))$ are high for all univariate series. In contrast, $k=8$ and $k=10$ correspond to pronounced outbreaks with small empirical probabilities.

\begin{table}
\caption{Expected values of a triavariate INAR(1) series with independent
Poisson innovations at the time of an outbreak of expected size $\kappa$ equal to $5, 8$ and $10$. 
Assumed parameter values are equal to $\mathbf{\theta}=(\alpha_{11},\alpha_{12},\alpha_{13},\alpha_{21},\alpha_{22},$ $\alpha_{23},\alpha_{31},\alpha_{32},\alpha_{33},\lambda_1,\lambda_2,\lambda_3)=(0.3, 0.1,0.2,0.2,0.4,0.2,0.3,0.2,0.2,1,1,1)$.}
 \label{tab:kappa-x} \centering
\begin{scriptsize}
\begin{tabular}{{c}{c}{c}{c}}
\hline
$\kappa$ & $E(X_{1t})$ & $E(X_{2t})$ & $E(X_{3t})$\\
\hline
5 & 7.9 & 8.7 & 8.3\\
8 & 10.9 & 11.7 & 11.3\\
10 & 12.9 & 13.7 & 13.3\\
\hline
\end{tabular}
\end{scriptsize}
\end{table}

\begin{table}
\caption{Empirical probabilities of the maximum value of each univariate series being greater than the corresponding expectation of the series at the time of an outbreak of expected size $\kappa$ equal to 5, 8 and 10. Data have been simulated with $n=200$ and $\mathbf{\theta}=(\alpha_{11},\alpha_{12},\alpha_{13},\alpha_{21},\alpha_{22},\alpha_{23},\alpha_{31},\alpha_{32},\alpha_{33},\lambda_1,\lambda_2,\lambda_3)= (0.3, 0.1,0.2,0.2,0.4,0.2,$ $0.3,0.2,0.2,1,1,1)$. The outbreak is assumed to happed at time $t=170$.}
 \label{tab:kappa-xmax} \centering
\begin{scriptsize}
\begin{tabular}{{c}{c}{c}{c}}
\hline
& \multicolumn{3}{c}{$P(\underset{t\neq 170}{\max}(x_{it})>E(X_{i,170}))$}\\
$\kappa$ & $i=1$ & $i=2$ & $i=3$\\
\hline
5 & 0.863 & 0.938 & 0.788 \\
8 & 0.072 & 0.171 & 0.068 \\
10 & 0.006 & 0.020 & 0.006\\
\hline
\end{tabular}
\end{scriptsize}
\end{table}

For each scenario ($\kappa=5, 8$ or $10$), we conducted $1000$
simulation replicates. In each replicate, a trivariate INAR(1)
model with independent Poisson innovations was fitted to the
set-up phase and the parameter estimates were used to compute
$90\%$, $95\%$ and $99\%$ upper prediction limits for the
monitoring phase. For comparison purposes, we also fitted three independent
INAR(1) models with Poisson innovations to the historical data and 
followed the same process for the computation of upper prediction
limits.

As evaluation measures we used the detection rate and weekly false alarm
rate based on a rule of $2/3$ that is, assuming that an alarm is
triggered if at least two out of the three series flagged an alarm
at the same point in time. The detection rate was computed as the
proportion of the 1000 replicates in which an alarm was triggered
at time $t=170$ while the weekly false alarm rate was defined as the number of cases
 in which an alarm was flagged at time $t\neq170$ divided by $1000\times 49$.
 Based on our simulations, we have also approximated the average run length (ARL) for different outbreak sizes $\kappa$ and different significance levels $\alpha$. In particular, for each univariate series we used the standard definition of ARL that is, we defined ARL$_i$, $i=1,2,3$ as the average number of points in the monitoring phase that precede the very first indication of a false alarm. Then, to get an overall measure of the performance of the suggested multivariate surveillance approach, we followed a conservative approach defining ARL=$\underset{i}{\min}$ARL$_{i}$.
However, it is important to note that this approximation and the related results should be treated with caution, first of all due to the limited number of simulations \cite{weiss2015}, secondly because we are handling multivariate count data through a multivariate surveillance approach, and thirdly because our decision on the occurrence of an outbreak is based on a $2/3$ rule rather than on modelling each series separately.

The estimated ARL$_i$'s and ARLs are summarized in Table~\ref{tab:arls}. The ARLs range from 12.6 to 13.1, 17.9 to 18.1 and 21.7 to 23.3 for $\alpha=10\%, 5\%$ and $1\%$ respectively when the multivariate approach is applied. Keeping in mind that the true outbreak has actually occurred at $t=170$ and that the monitoring phase is the period $t=150,\ldots, 200$, we conclude that if a false alarm is triggered, this is expected to happen around a week earlier than the true outbreak  when $\alpha=10\%$, around the time of the true outbreak when $\alpha=5\%$ and a bit later than the time of the true outbreak when $\alpha=1\%$. Fitting three independent INAR(1) models to the data results in generally lower ARLs that range between 10.1 and 10.3 when $\alpha=10\%$, between 14.1 and 15.1 when $\alpha=5\%$ and between 19.5 and 21.6 when $\alpha=1\%$. The consistently higher ARLs obtained by the multivariate approach indicate its superiority over the univariate modeling approach in terms of this specific evaluation measure.

\begin{table}
\caption{Average run lengths for the three series (ARL$_i$'s) and overall (ARLs), for
different outbreak sizes $\kappa$ and different significance
levels $\alpha$.} \label{tab:arls} \centering
\begin{scriptsize}
\begin{tabular}{{c}{l}{c}{c}{c}{c}{c}{c}{c}{c}}
\hline
& & \multicolumn{4}{c}{trivariate INAR(1)} & \multicolumn{4}{c}{independent INAR(1)} \\
Outbreak size & Sign. level &  ARL$_1$ & ARL$_2$ & ARL$_3$ & ARL & ARL$_1$ & ARL$_2$ & ARL$_3$ & ARL \\
$\kappa=5$ & $\alpha=10\%$ & 13.6 & 13.1 & 13.1 & 13.1 & 14.8 & 10.3 & 11.1 & 10.3\\
& $\alpha=5\%$ & 18.4 & 18.7 & 17.9 & 17.9 & 19.0 & 15.1 & 16.0 & 15.1\\
& $\alpha=1\%$ & 22.2 & 23.9 & 21.7 & 21.7 & 22.3 & 21.6 & 21.6 & 21.6\\
\hline 
$\kappa=8$ & $\alpha=10\%$ & 13.0 & 13.1 & 12.6 & 12.6 & 14.4 & 10.3 & 11.2 & 10.3\\
& $\alpha=5\%$ & 18.5 & 18.5 & 18.1 & 18.1 & 18.4 & 15.0 & 15.8 & 15.0\\
& $\alpha=1\%$ & 22.5 & 23.4 & 22.4 & 22.4 & 21.4 & 21.0 & 21.1 & 21.0\\
\hline
$\kappa=10$ & $\alpha=10\%$ & 13.8 & 13.0 & 13.5 & 13.0 & 14.4 & 10.1 & 11.2 & 10.1\\
& $\alpha=5\%$ & 18.1 & 18.7 & 18.8 & 18.1 & 18.8 & 14.1 & 15.3 & 14.1\\
& $\alpha=1\%$ & 26.6 & 23.3 & 24.0 & 23.3 & 22.0 & 19.5 & 20.6 & 19.5\\
\hline
\end{tabular}
\end{scriptsize}
\end{table}

The estimated detection rates and false alarm rates are summarized in Table~\ref{tab:sims}. As
expected, the larger the size of the outbreak is, the higher the
achieved detection rate. This conclusion holds for both the trivariate and the indepedent INAR(1) modeling approaches that are equivalently effective in terms of the estimated detection rates. However, the multivariate approach has an obvious superiority in terms of the false alarm rates that are consistently lower than the corresponding false alarms rates achieved for all $\kappa$'s  and $\alpha$'s under the univariate approach. The outperformance of the multivariate approach in terms of false alarm rates is not surprising since the independent INAR(1) models ignore the cross-correlation between the series resulting in narrower 
prediction intervals and thus increasing the number of false alarms.

Focusing on the multivariate approach, it is evident that the false alarm rates are generally low without any particular pattern with regard to the outbreak size.
 Regarding the role of the
significance level $\alpha$, we observe that decreasing $\alpha$
results in lowering both the detection rates and false alarm
rates. The degree of reduction depends however on the true oubreak
size. In particular, the conservative $\alpha=1\%$ proves to be
too strict for $\kappa=5$ as it achieves a detection rate of
around $55\%$ contrary to $\alpha=5\%$ or $10\%$ that achieve
detection rates of $80\%$ and $89\%$ respectively. However,
the detection rates achieved at different significance levels
improve considerably for larger outbreak sizes even reaching
$99.8\%$ for $\kappa=10$ and $\alpha\geq 5\%$. For $\alpha=1\%$ the
corresponding detection rate is equal to $98.5\%$ but with a false
alarm rate of $0.03\%$ that is much smaller than those
corresponding to $\alpha=5\%$ or $10\%$ ($0.40\%$ and $1.44\%$
respectively). Conclusively, for the choice of the significance
level to be used for outbreak detection purposes, one should
consider the outbreak size that he or she would like to detect and
 the false alarm rate that is ready to accept. In the following
section we use the conservative $\alpha=1\%$.

\begin{table}
\caption{Detection rates (DR) and false alarm rates (FAR) for
different outbreak sizes $\kappa$ and different significance
levels $\alpha$. The reported numbers have been multiplied by
100.} \label{tab:sims} \centering
\begin{scriptsize}
\begin{tabular}{{c}{l}{c}{c}{c}{c}}
\hline
& & \multicolumn{2}{c}{trivariate INAR(1)} & \multicolumn{2}{c}{independent INAR(1)}\\
Outbreak size & Sign. level & DR & FAR & DR & FAR\\
\hline
$\kappa=5$ & $\alpha=10\%$ & 89.0 & 1.33 & 88.6 & 2.95\\
& $\alpha=5\%$ & 80.1 & 0.34 & 78.4 & 0.99\\
& $\alpha=1\%$ & 55.1 & 0.01 & 49.5 & 0.10\\
\hline
$\kappa=8$ & $\alpha=10\%$ & 99.4 & 1.30 & 99.3 & 3.80\\
& $\alpha=5\%$ & 98.7 & 0.32 &  98.0 & 1.51\\
& $\alpha=1\%$ & 93.4 & 0.01 & 91.4 & 0.23\\
\hline 
$\kappa=10$ & $\alpha=10\%$ & 99.8 & 1.44 & 99.9 & 4.22\\
& $\alpha=5\%$ & 99.8 & 0.40 & 99.7 & 1.93\\
& $\alpha=1\%$ & 98.5 & 0.03 & 98.1 & 0.36\\
\hline
\end{tabular}
\end{scriptsize}
\end{table}

\section{Application using Syndromic data}

 Among various aspects of health surveillance, syndromic surveillance is considered as an important tool since it is based on symptoms rather than diagnosis and hence it
can create alerts faster.  For example, syndromic surveillance systems for detection of biologic terrorism after the terrorist attack of September 11, 2001 have been launched in New York city \citep{das2003}. In addition, during large athletic events such surveillance systems can be useful to quickly detect threats for the public health, and they have been used in winter Olympic Games in Salt Lake City 2002 \citep{gesterland2003, mundorff2004} and Athens 2004 Olympic Games \citep{dafni2004}.
Syndromic surveillance data are by nature low count data, especially if they refer to incidences of diseases and symptoms that are not so common. In such cases, the usual normal approximation is not appropriate and the data should rather be treated as discrete-valued time series. Moreover, when the collected data involve several related variables, this brings forward the need to consider multivariate surveillance techniques.

The data considered here is part of the syndromic surveillance data collected during Athens 2004 Olympic Games. The full database consists of eleven different syndromes recorded  since July 2002 in emergency departments of major hospitals in the Greater Athens area (drop-in syndromic surveillance) \citep{dafni2004}. For the purpose of the current analysis we consider three distinct syndromes recorded in a specific hospital that are significantly correlated to each other (cross-correlations ranging from $0.31$ to $0.48$). In particular, we consider respiratory infection with fever, febrile illness with rash and other syndrome with potential interest for public health.
The latter is a general category including all symptoms that could not be classified in any of the other prespecified categories.

\begin{figure}
\centering
\includegraphics[scale=0.6]{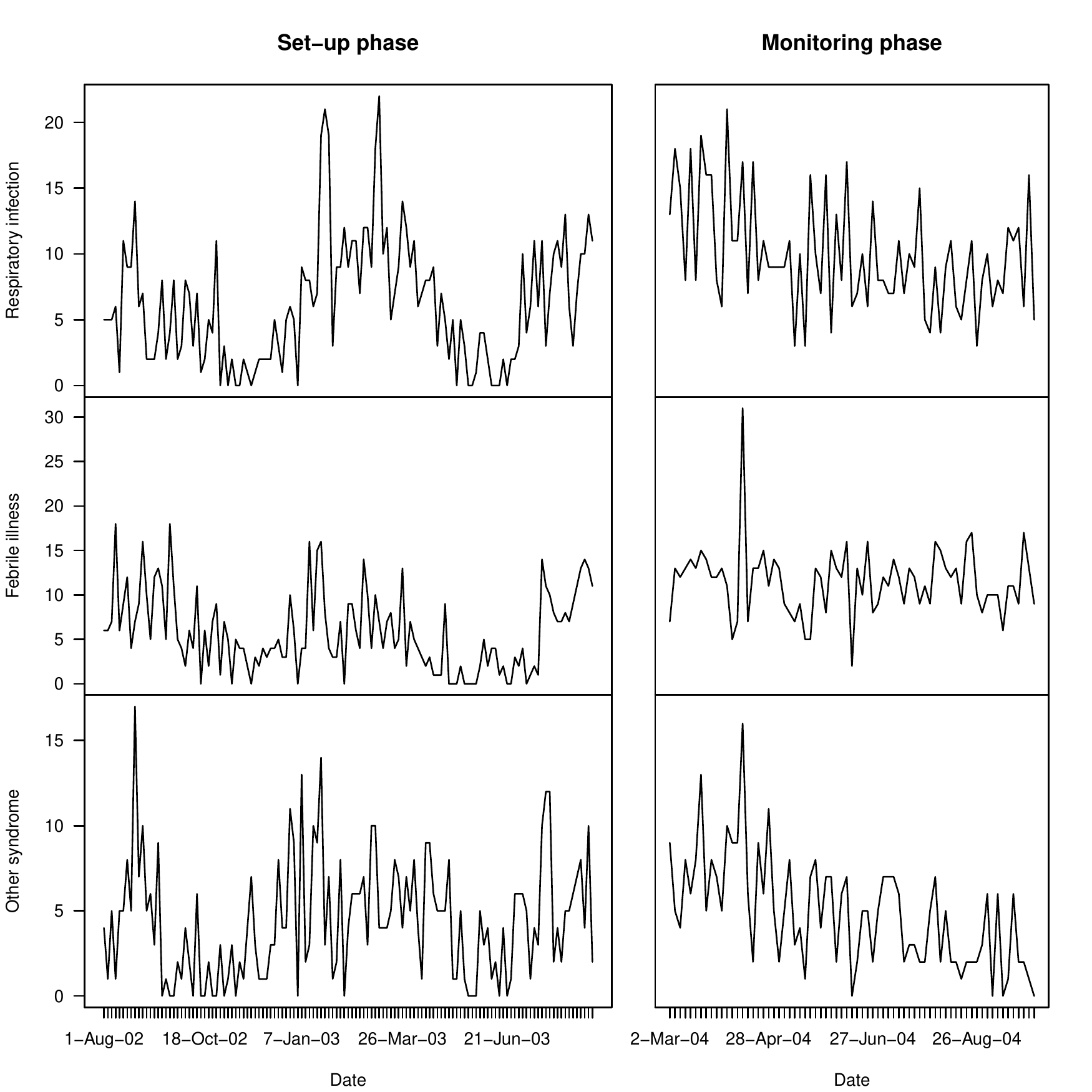}
\caption{Time series plots of the data}
\label{app-fig1}
\end{figure}

\begin{figure}[htbp]
\centering
\begin{tabular}{cc}
\includegraphics[scale=0.38]{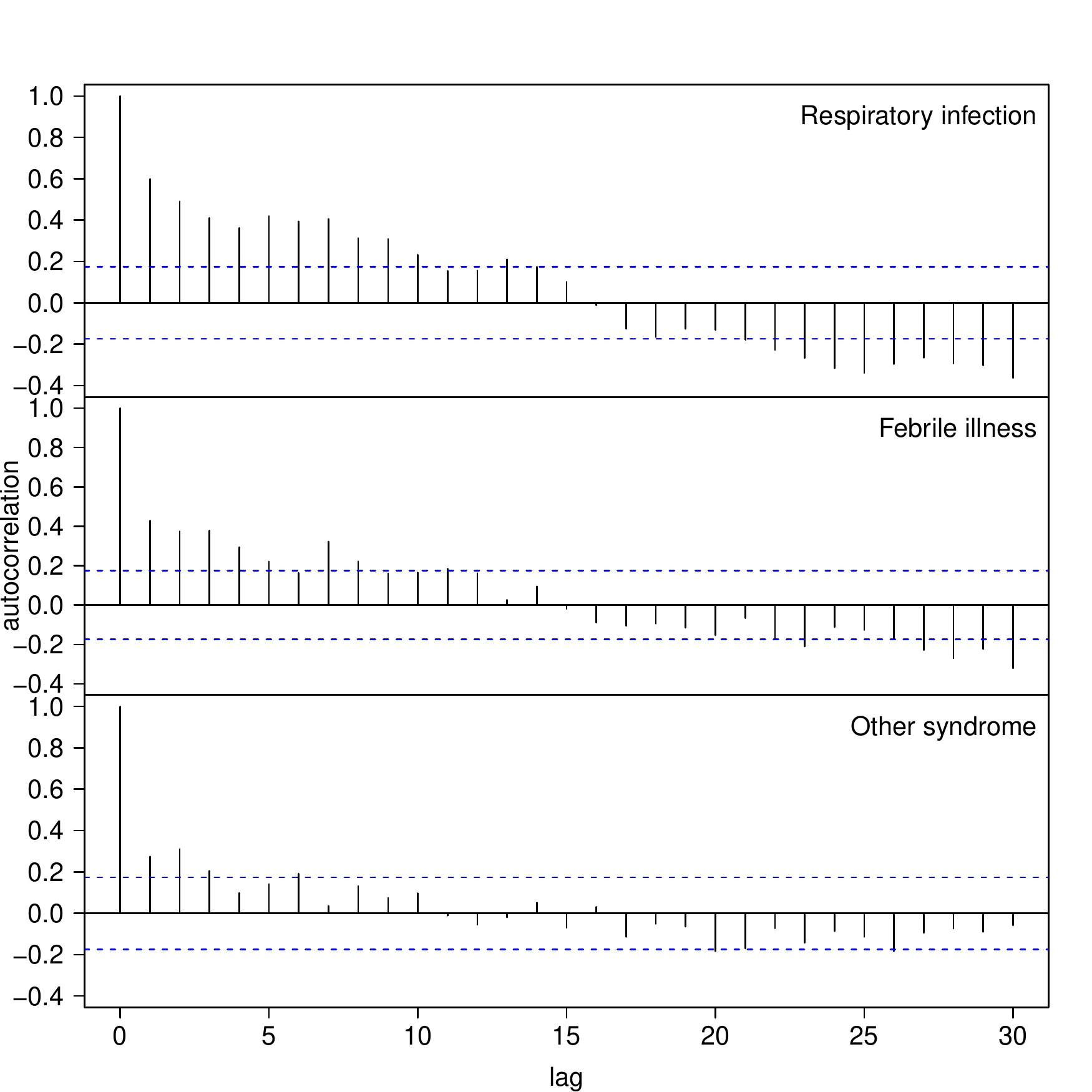} & \hfill \includegraphics[scale=0.38]{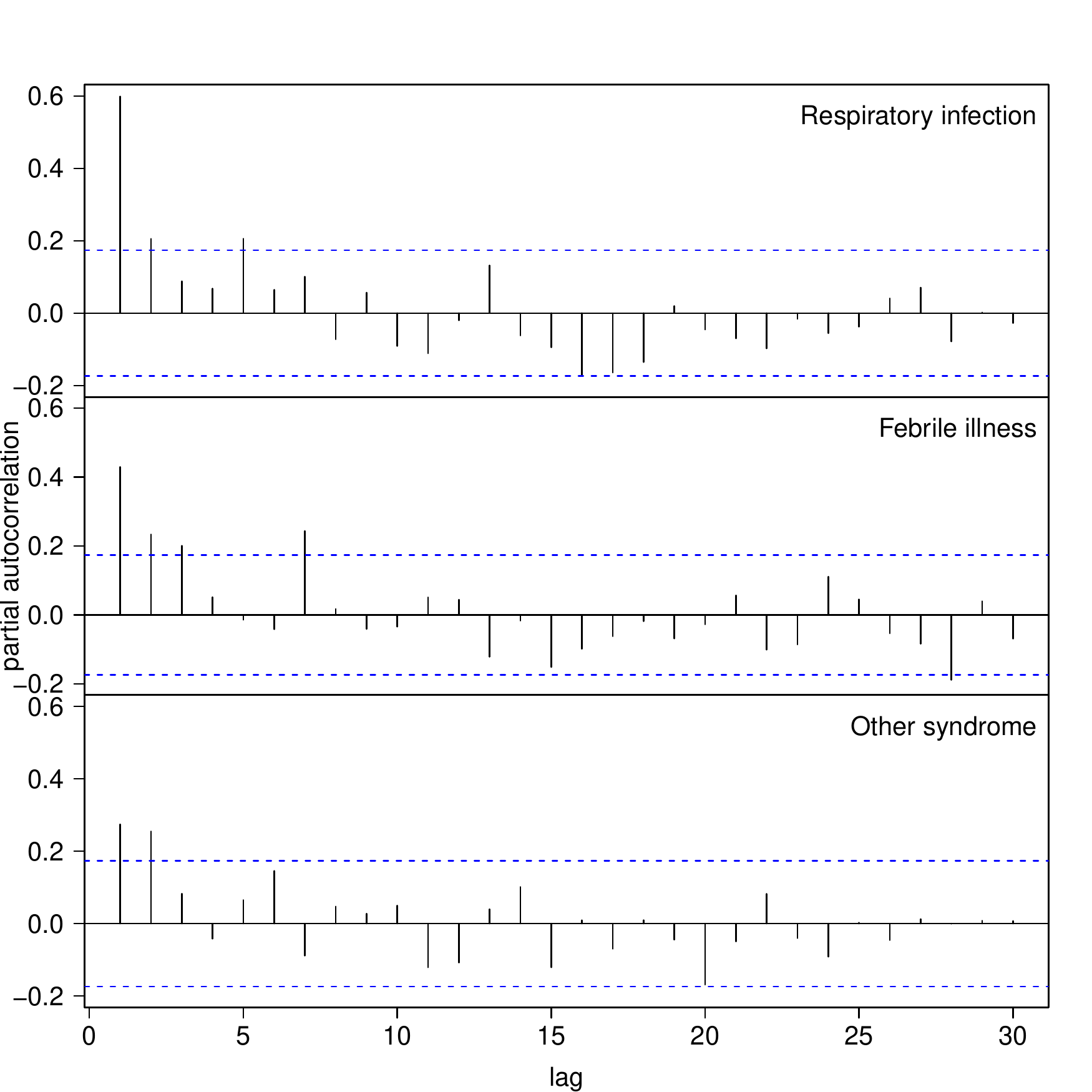}
\end{tabular}
\caption{Plots of the autocorrelations (left panel) and partial autocorrelations (right panel) of the historical data}
\label{app-fig2}
\end{figure}

Our monitoring period starts on March 2, 2004 and ends on September 28, 2004 while the period between August 1, 2002 and August 29, 2003 is considered as the set-up phase. During both periods syndromes were recorded every three days so that the historical and surveillance data consist of $t_0 =127$  and $t_1 = 71$ observations respectively.
The time series plots of the three series during the set-up and monitoring phases are included in Figure~\ref{app-fig1}. Table~\ref{app-tab1} summarizes basic descriptive statistics.
The plots of autocorrelations and partial autocorrelations of the three series during the set-up phase are shown in Figure~\ref{app-fig2}. The exponentially decaying autocorrelation functions indicate the appropriateness of an AR-type modeling approach whilst the partial autocorrelation functions suggest an order of dependence around one or two.

In the following we apply the approach of Section~\ref{sec-survmodel}, i.e.  we fit a trivariate INAR(1) model with indepedent Poisson innovations for modeling and prediction using the historical syndromic surveillance data. To account for regressors usually related to infectious disease data we express the expectation of the innovation series as function of the available covariate information, i.e. $E(\epsilon_{it}) = \exp( \mathbf{z}_t'\boldsymbol{\beta})$, $i=1,2,3$, where $\mathbf{z}_t$ as a vector of covariates with associated regression parameters $\boldsymbol{\beta}$ \citep{pedeli2013b}.
As candidate covariates we consider terms for  seasonality and a binary indicator for the day of the week on which the recording of syndromes was implemented (weekdays vs. weekends). We don't consider time trends since  Figure~\ref{app-fig1} does not suggest the presence of any trend in our data.
Therefore, each marginal series is modeled as
$X_{it}=\sum_{j=1}^{3}\alpha_{ij}\circ X_{j,t-1} + \epsilon_{it}, i=1, 2, 3$, where $\epsilon_{it}$ are independent Poisson random variables with mean
\begin{equation}
\label{eq:regmean}
E(\epsilon_{it})=\exp\left\{\beta_{i0}+\beta_{i1}\mbox{Weekday}+\beta_{i2}\cos{\left(\frac{2\pi t}{122}\right)} + \beta_{i3}\sin{\left(\frac{2\pi t}{122}\right)}\right\}
\end{equation}
for $t=1, \ldots, t_0$. Note that in the trigonometric terms that have been employed to capture seasonal patterns, we consider a seasonal period equal to $122$ because our data are recorded in three-days intervals.
 For comparison purposes we also employ a univariate surveillance approach based on fitting three indepedent INAR(1) regression models with Poisson innovations. Covariate information is incorporated in the univariate models in the same way, i.e. through (\ref{eq:regmean}).
With both approaches, the marginal one-step-ahead predictive distributions are used for the construction of $(1-\alpha)\%$ prediction intervals, the upper bounds of which serve as thresholds for outbreak detection. We assume a component-wise type I error rate of $\alpha=0.01$ and for the overall alarm we set a rule of $2/3$ that is an alarm is triggered if at least two out of the three series flag an alarm at the same point in time.

\begin{center}
\begin{figure}[htbp]
\centering
\begin{tabular}{cc}
\includegraphics[scale=0.38]{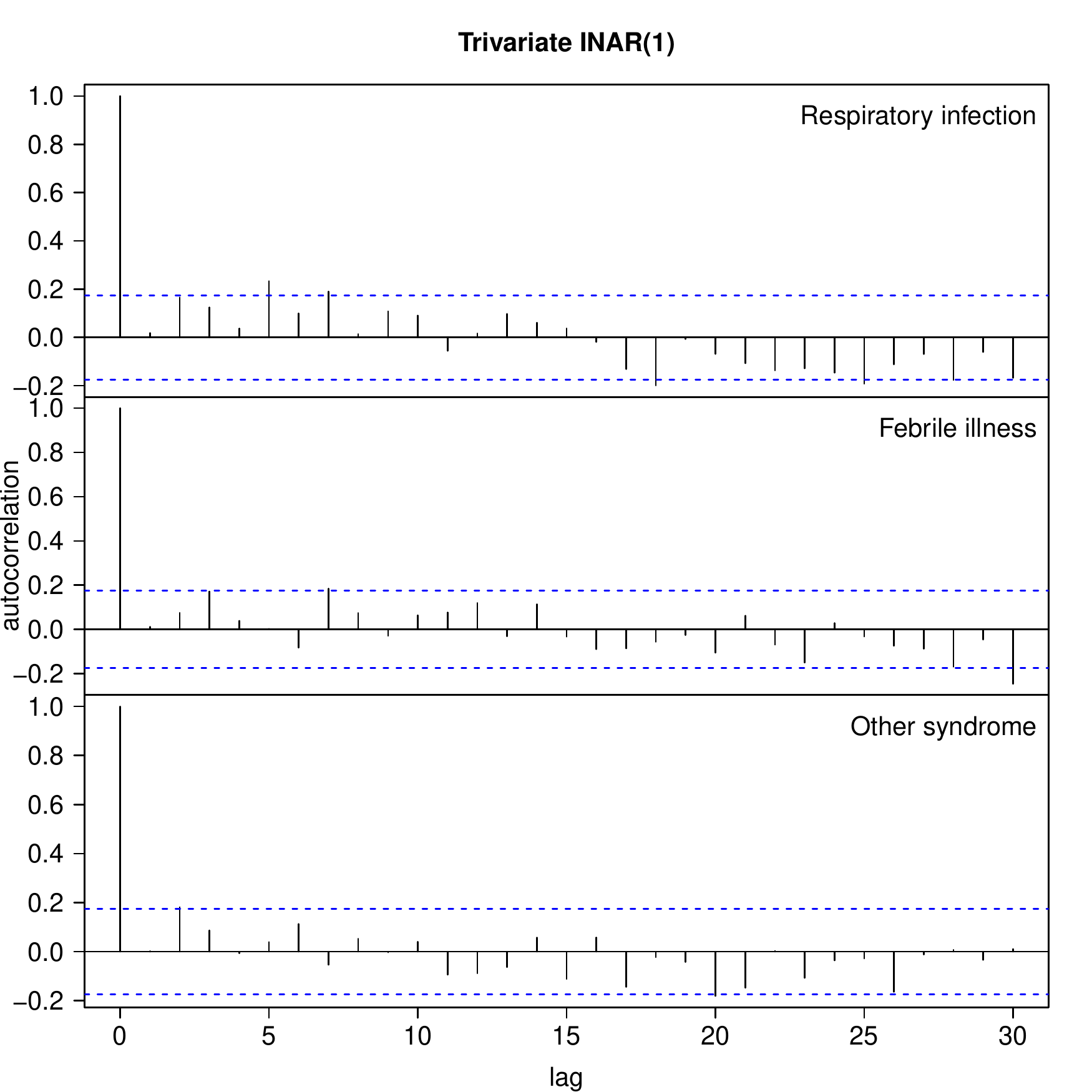} & \hspace{-18pt} \includegraphics[scale=0.38]{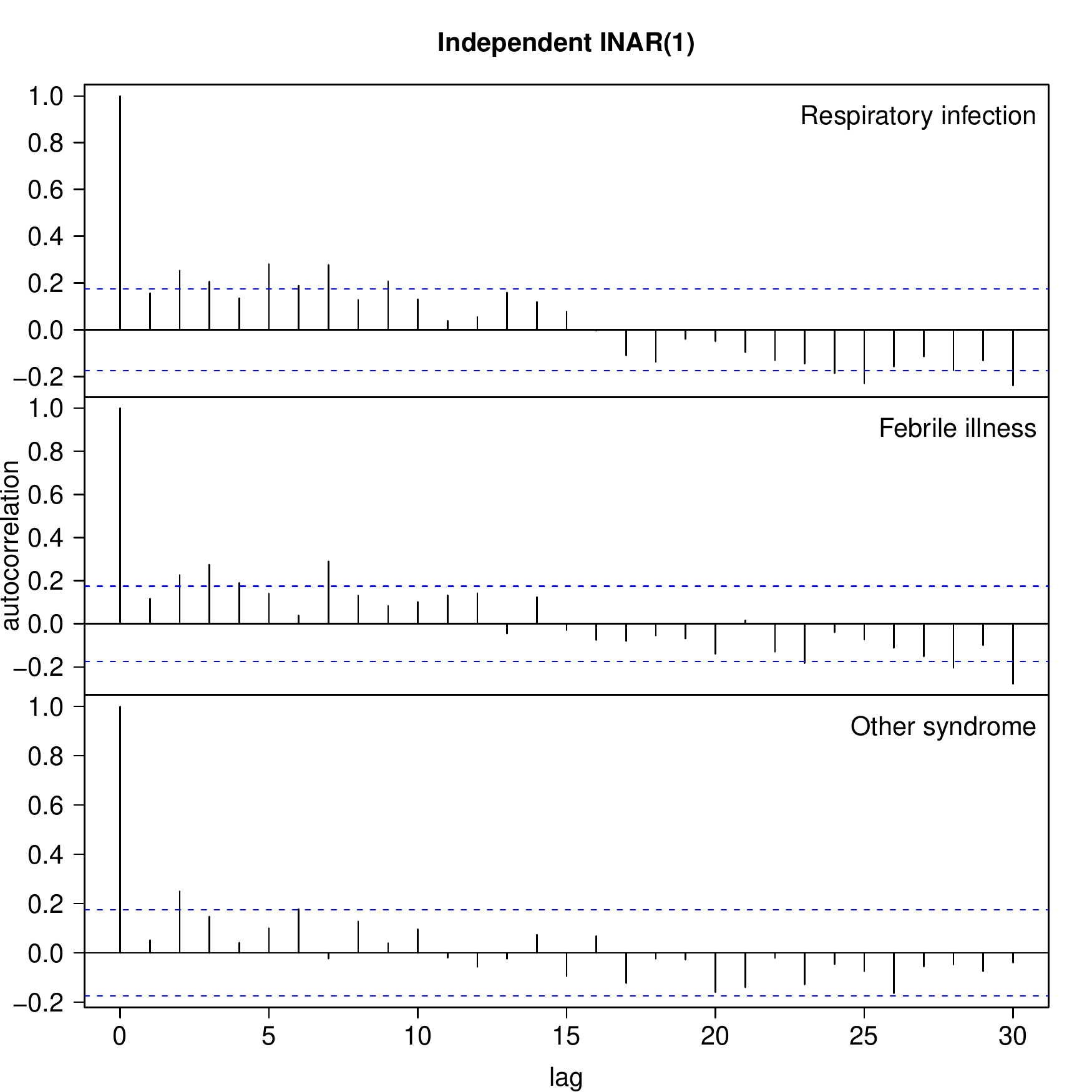}
\end{tabular}
\caption{Plots of the autocorrelations of the residuals obtained by the trivariate INAR(1) (left panel) and the independent INAR(1) (right panel) regression models.}
\label{app-fig3}
\end{figure}
\end{center}

\begin{center}
\begin{figure}[htbp]
\centering
\begin{tabular}{cc}
\includegraphics[scale=0.38]{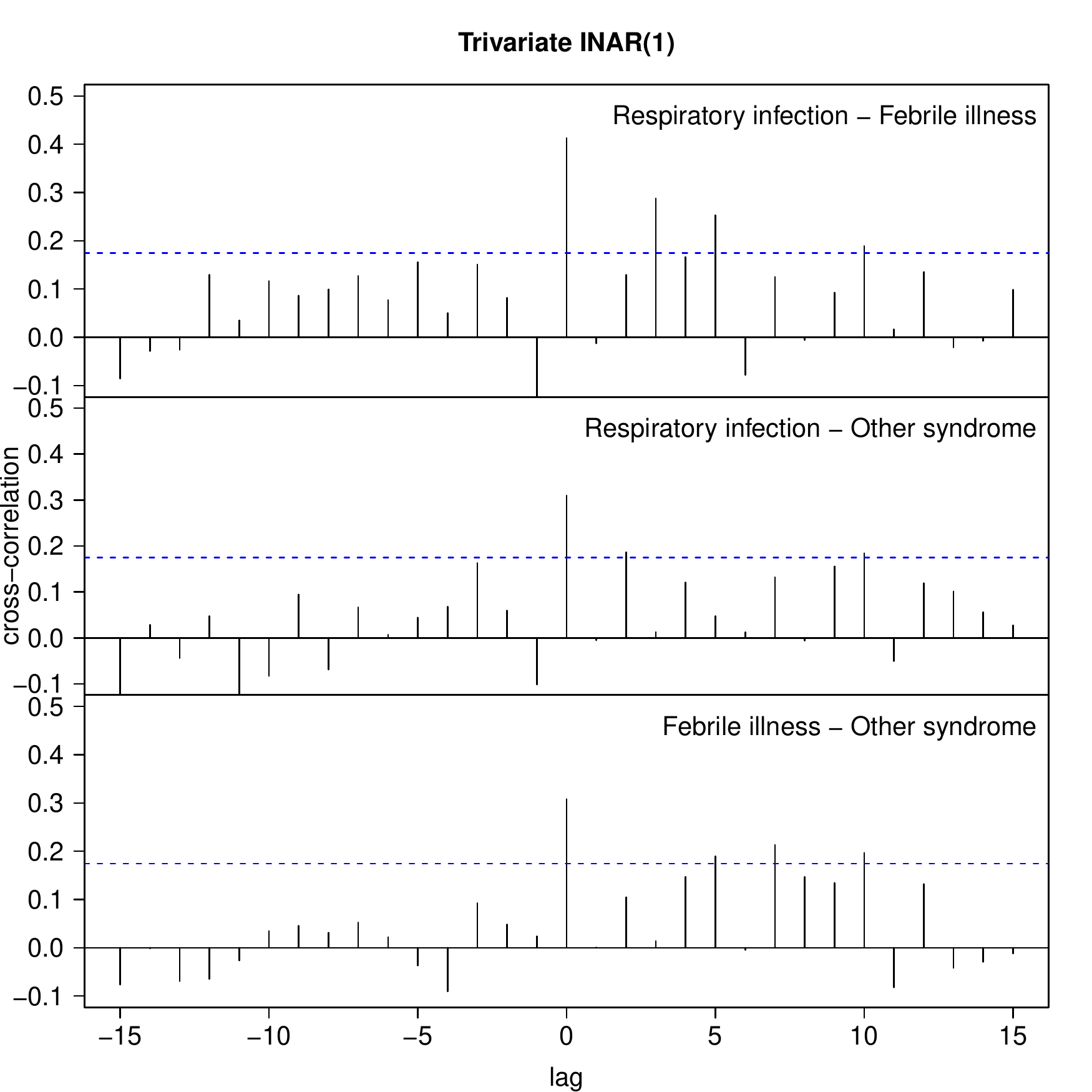} & \hspace{-18pt} \includegraphics[scale=0.38]{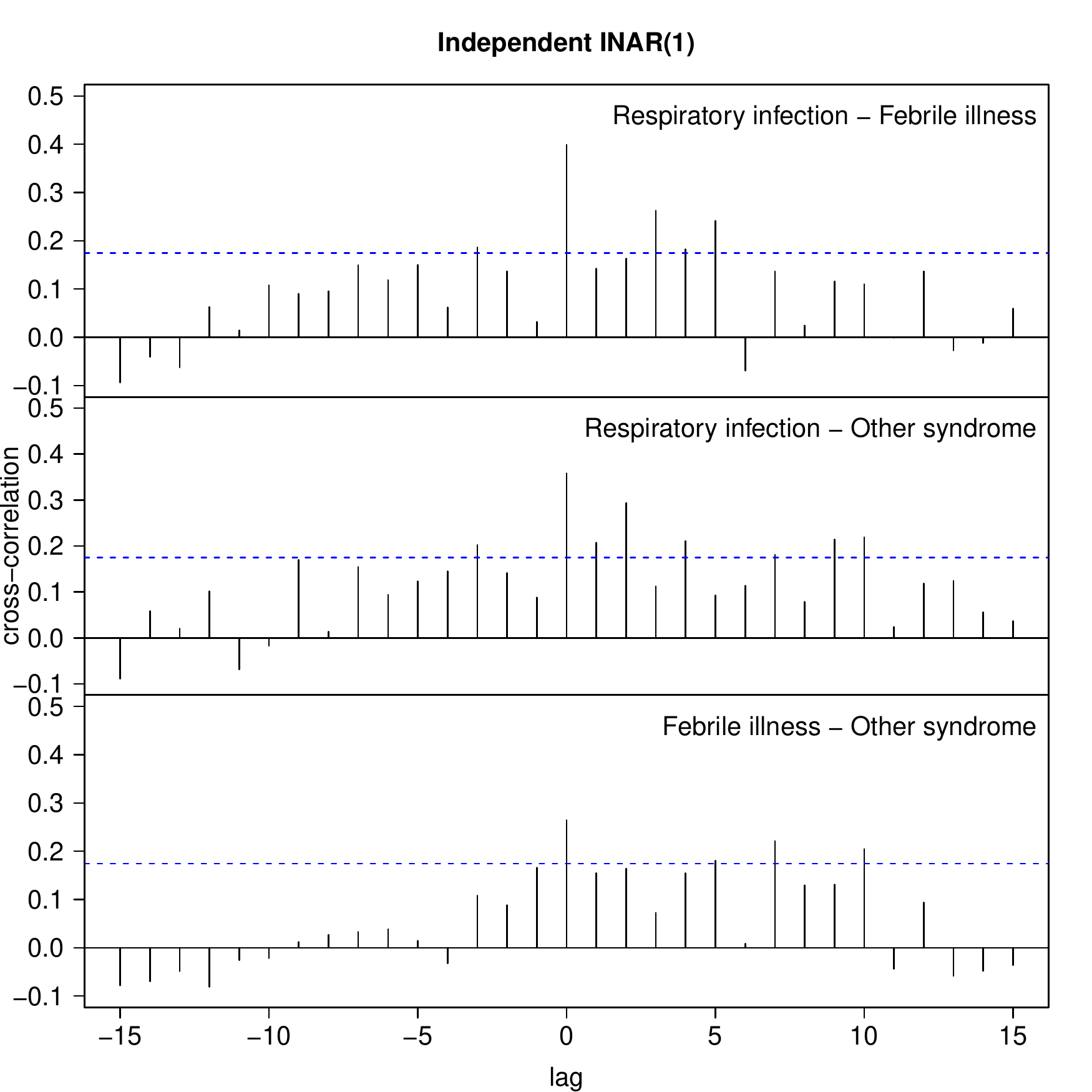}
\end{tabular}
\caption{Plots of the cross-correlations of the residuals obtained by the trivariate INAR(1) (left panel) and the independent INAR(1) (right panel) regression models.}
\label{app-fig5}
\end{figure}
\end{center}

The parameter estimates and corresponding standard errors obtained with the two modeling approaches are summarized in Table~\ref{app-tab2}. Results indicate significant
first-order autocorrelations under both fittings. The cross-correlation parameters estimated by the trivariate INAR(1) model are also significant indicating the appropriateness of the multivariate approach. Figure~\ref{app-fig3} shows the correlograms of the residuals obtained by the two modeling approaches. Obviously, the trivariate INAR(1) regression model can effectively capture significant autocorrelations at almost all lags while the three univariate INAR(1) models are less effective in accounting for autocorrelations greater than one. However, Figure~\ref{app-fig5} reveals some remaining cross-correlations with both approaches although such cross-correlations are more persistent with the univariate INAR(1) models.

\begin{center}
\begin{figure}[]
\begin{tabular}{cc}
\includegraphics[scale=0.38]{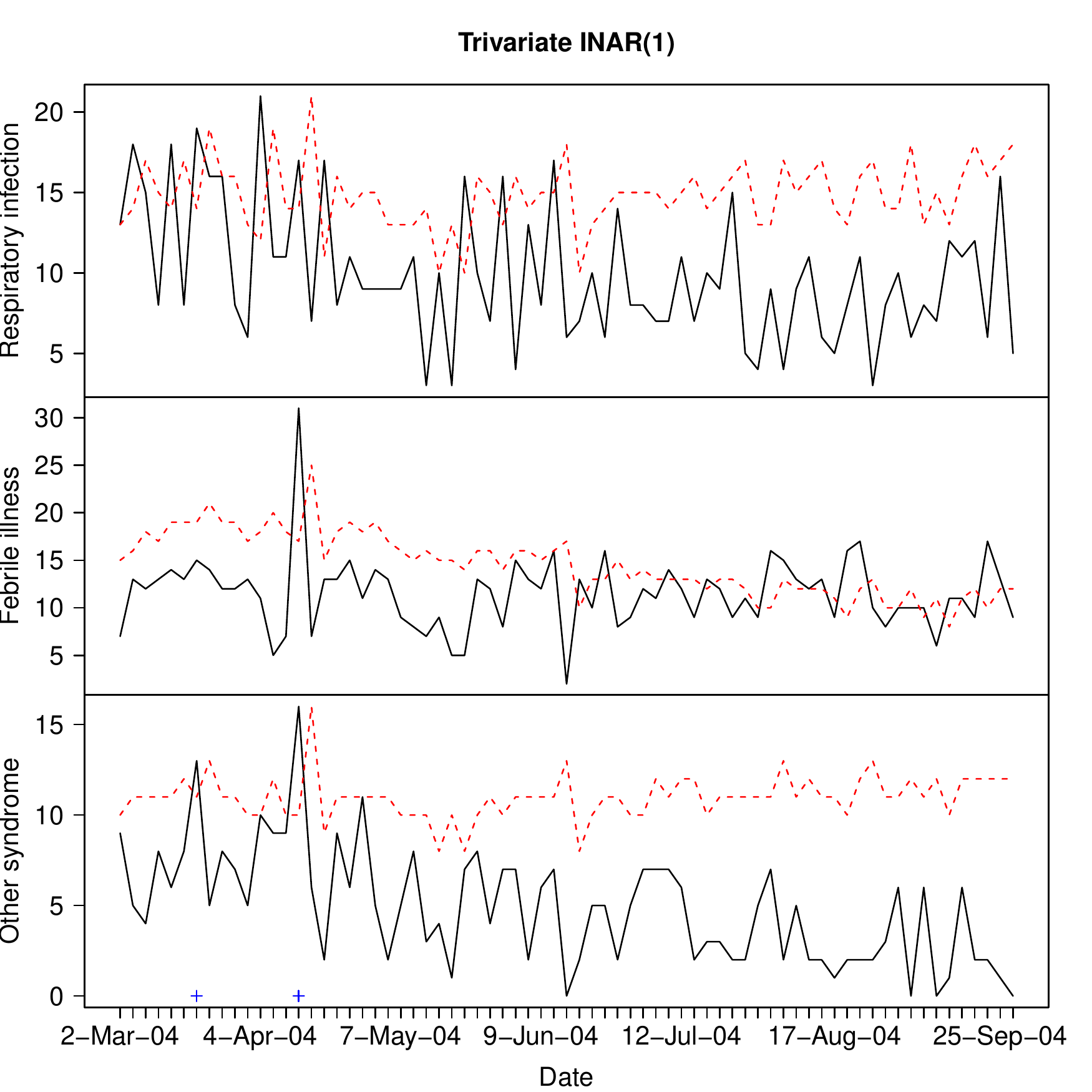} & \hspace{-18pt} \includegraphics[scale=0.38]{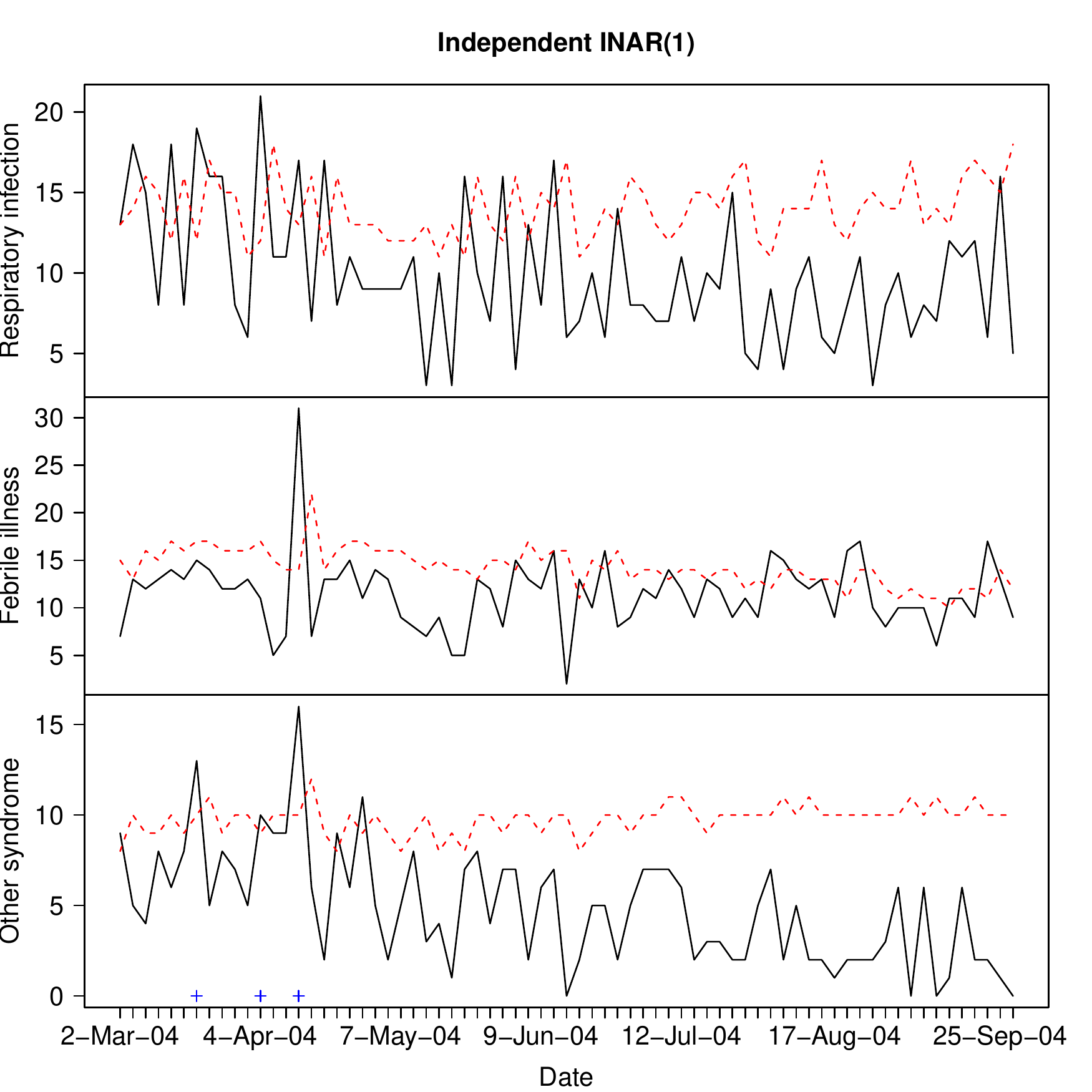}
\end{tabular}
\caption{ Surveillance plots as obtained after fitting a trivariate INAR(1) regression model with independent Poisson
innovations (left panel) or three independent Poisson INAR(1) regression models (right panel) to the historical data. Statistical alarms (blue crosses) are raised when at least two series exceed the upper bounds of the corresponding 99\%
prediction intervals (red dashed lines).}
\label{app-fig4}
\end{figure}
\end{center}

The surveillance plots obtained under the two models are shown in
Figure~\ref{app-fig4}. Red dashed lines represent the upper bounds
of the corresponding 99\% prediction intervals while blue crosses
indicate the time points at which an overall alarm is raised. The
two alarms signalled with the trivariate INAR(1) fitting are also
trigerred when three independent INAR(1) models are fitted to the
historical data but the later approach also gives an additional
alarm.

\section{Discussion}

As the amount of available data increases, multivariate
surveillance scenarios become more and more plausible. Aiming to
contribute in this developing area, we suggest a multivariate
INAR(1) approach, suitable for joint modeling of multivariate
surveillance data. The introduced model admits a realistic
epidemiological interpretation with a clear distinction between
the epidemic and endemic components and accounts for
overdispersion that is typical with surveillance data.  Even
though emphasis has been put on the case of independent Poisson
innovations, other discrete distributions, as e.g. the negative
binomial, can also be considered instead.

In this paper we provided a generic framework for using models
suitable for multivariate counts time series for surveillance
purposes. A series of interesting points that refer to the context
of surveillance can be further exploited, as for example updating
the data basis for the model fit in a regular basis and keep the
newest observations only for building the model (see
\cite{noufaily2013improved}) or downweight past outbreaks by
suitable adjustments (e.g. in \cite{noufaily2013improved} Anscombe
residuals were used). Of course note that, because of the
discreteness of the data, it is not obvious how methods suitable
for continuous and univariate outcomes translate to our case and
this is an interesting topic for further research. 

A final
comment relates to the parametric assumptions made in this paper.
While our model can capture small to moderate overdispersion one
may alter the assumption about the innovations to allow for larger
overdispersion. Also note the notion of multivariate
overdispersion
 discussed in \cite{kokonendji2018fisher} which can be also a vehicle for building more flexible models
and examine their properties.

\begin{table}
\caption{Mean, variance and coefficient of variation (CV) for the three syndromes during the set-up and monitoring phases. }
\label{app-tab1}
\centering
\begin{scriptsize}
\begin{tabular}{{l}{c}{c}{c}|{c}{c}{c}}
\hline
& \multicolumn{3}{c}{Set-up phase} & \multicolumn{3}{c}{Monitoring phase}\\
\hline
  & Mean& Variance& CV & Mean & Variance & CV\\
\hline
Respiratory infection & 6.17 & 22.40 & 76.7\% &  \multicolumn{1}{r}{9.90} & 18.46 & 43.4\%\\
Febrile illness & 5.76 & 19.96 & 77.5\% & 11.48 & 15.37 & 34.2\%\\
Other syndrome & 4.39 & 12.34 & 79.9\% & \multicolumn{1}{r}{4.89} & 10.36 & 65.9\%\\
\hline
\end{tabular}
\end{scriptsize}
\end{table}

\begin{table}
\caption{Maximum likelihood estimates (standard errors) from
fitting three independent Poisson INAR(1)  or a trivariate INAR(1) regression model with independent Poisson
innovations to the historical data.}
\label{app-tab2}
\centering
\begin{scriptsize}
\begin{tabular}{{c}{r}{r}|{c}{r}{r}}
\hline
correlation parameters & trivariate INAR(1) & independent INAR(1) & regression parameters& trivariate INAR(1) & indepedent INAR(1) \\
\hline
 $\hat{\alpha}_{11}$ & 0.329 (0.044) & 0.393 (0.039) & $\hat\beta_{10}$ &  1.190 (0.153) & 1.506 (0.099)\\
$\hat{\alpha}_{12}$ &  0.126 (0.043) & - & $\hat\beta_{11}$ & -0.255 (0.145) & -0.278 (0.110)\\
$\hat{\alpha}_{13}$ &  0.134 (0.054) & - & $\hat\beta_{12}$ & -0.359 (0.118) & -0.222 (0.078)\\
$\hat{\alpha}_{21}$ & 0.160 (0.040) & - & $\hat\beta_{13}$ &  -0.218 (0.098) & -0.140 (0.073)\\
$\hat{\alpha}_{22}$ & 0.177 (0.045) & 0.263 (0.041) & $\hat\beta_{20}$ &  1.197 (0.135) & 1.496 (0.096)\\
$\hat{\alpha}_{23}$ &  0.141 (0.048) & - & $\hat\beta_{21}$ & -0.267 (0.133) & -0.118 (0.102)\\
$\hat{\alpha}_{31}$ & 0.062 (0.039) & - & $\hat\beta_{22}$ & 0.411 (0.121) & 0.156 (0.070)\\
$\hat{\alpha}_{32}$ & 0.108 (0.039) & - & $\hat\beta_{23}$ & 0.548 (0.110) & 0.296 (0.068)\\
$\hat{\alpha}_{33}$ & 0.131 (0.047)& 0.179 (0.045) & $\hat\beta_{30}$ & 0.990 (0.155) & 1.246 (0.109)\\
 & & & $\hat\beta_{31}$ & 0.047 (0.142) & 0.046 (0.113)\\
 & & & $\hat\beta_{32}$ & -0.174 (0.099) & -0.112 (0.072)\\
 & & & $\hat\beta_{33}$ & -0.198 (0.090) & -0.146 (0.071)\\
\hline
\end{tabular}
\end{scriptsize}
\end{table}

\section*{Acknowledgements}
This project has received funding from the Athens University of
Economics,  Action II Funding \& Research Funding Program no. 2938-01 and from the European Union's Horizon
2020 research and innovation programme under the Marie
Sk{\l}odowska-Curie grant agreement no. 699980.



\bibliographystyle{chicago}

\bibliography{Surveillance_biblio}

\end{document}